\documentclass[%
 superscriptaddress,
singlecolumn,
 amsmath,amssymb,
 aps,prl
]{revtex4-2}

\usepackage{graphicx}% Include figure files
\usepackage{dcolumn}% Align table columns on decimal point
\usepackage{xcolor}
\usepackage{bm}% bold math
\usepackage[hidelinks]{hyperref}% add hypertext capabilities

\usepackage{siunitx} % SI unit formatting
\usepackage{tabularx} % more control over tables
\usepackage{siunitx}
\usepackage{makecell}
\usepackage[T1]{fontenc}
\begin{document}

%\preprint{APS/123-QED}

\title{Further evidence for shape coexistence in $^{79}$Zn$^{m}$ near doubly-magic $^{78}$Ni}

\author{L.~Nies}
\email{Lukas.Nies@cern.ch}
\affiliation{European Organization for Nuclear Research (CERN), Meyrin, 1211 Geneva, Switzerland}
\affiliation{Institut für Physik, Universität Greifswald, 17487 Greifswald, Germany}

\author{L.~Canete}
\affiliation{University of Jyvaskyla, Department of Physics, Accelerator laboratory, P.O. Box 35(YFL) FI-40014 University of Jyvaskyla, Finland}%
\affiliation{Department of Physics, University of Surrey, Guildford, GU2 7X5, United Kingdom}

\author{D.~D.~Dao}
\affiliation{Universit\'e de Strasbourg, CNRS, IPHC UMR 7178, F-67000 Strasbourg, France}

\author{S.~Giraud}
\altaffiliation{Present address: Facility for Rare Isotope Beams, Michigan State University, East Lansing, Michigan 48824, USA}
\affiliation{GANIL, Bd Henri Becquerel, BP 55027, F-14076 Caen Cedex 5, France}

\author{A.~Kankainen}
\email{anu.kankainen@jyu.fi}
\affiliation{University of Jyvaskyla, Department of Physics, Accelerator laboratory, P.O. Box 35(YFL) FI-40014 University of Jyvaskyla, Finland}

\author{D.~Lunney}
\affiliation{Université Paris-Saclay, CNRS/IN2P3, IJCLab, 91405 Orsay, France}

\author{F.~Nowacki}
\affiliation{Universit\'e de Strasbourg, CNRS, IPHC UMR 7178, F-67000 Strasbourg, France}

\author{B.~Bastin}
\affiliation{GANIL, Bd Henri Becquerel, BP 55027, F-14076 Caen Cedex 5, France}

\author{M.~Stryjczyk}
\affiliation{University of Jyvaskyla, Department of Physics, Accelerator laboratory, P.O. Box 35(YFL) FI-40014 University of Jyvaskyla, Finland}

\author{P.~Ascher} 
\affiliation{Universit\'e de Bordeaux, CNRS/IN2P3—Universit\'e, CNRS/IN2P3, LP2I Bordeaux, UMR 5797, F-33170 Gradignan, France}

\author{K.~Blaum}
\affiliation{Max-Planck-Institut für Kernphysik, 69117 Heidelberg, Germany}

\author{R.~B.~Cakirli}
\affiliation{Department of Physics, Istanbul University, Istanbul 34134, Turkey}

\author{T.~Eronen} 
\affiliation{University of Jyvaskyla, Department of Physics, Accelerator laboratory, P.O. Box 35(YFL) FI-40014 University of Jyvaskyla, Finland}

\author{P.~Fischer}
\affiliation{Institut für Physik, Universität Greifswald, 17487 Greifswald, Germany}

\author{M.~Flayol}
\affiliation{Universit\'e de Bordeaux, CNRS/IN2P3—Universit\'e, CNRS/IN2P3, LP2I Bordeaux, UMR 5797, F-33170 Gradignan, France}

\author{V.~Girard~Alcindor} 
\affiliation{GANIL, Bd Henri Becquerel, BP 55027, F-14076 Caen Cedex 5, France}

\author{A.~Herlert}
\affiliation{FAIR GmbH, Planckstraße 1, 64291 Darmstadt, Germany}

\author{A.~Jokinen} 
\affiliation{University of Jyvaskyla, Department of Physics, Accelerator laboratory, P.O. Box 35(YFL) FI-40014 University of Jyvaskyla, Finland}

\author{A.~Khanam}
\affiliation{University of Jyvaskyla, Department of Physics, Accelerator laboratory, P.O. Box 35(YFL) FI-40014 University of Jyvaskyla, Finland}
\affiliation{Department of Applied Physics, Aalto University, P.O. Box 15100, FI-00076 Aalto, Finland}
\affiliation{Department of Physics, University of Helsinki, P.O. Box 43, FI-00014 Helsinki, Finland}

\author{U.~Köster}
\affiliation{European Organization for Nuclear Research (CERN), Meyrin, 1211 Geneva, Switzerland}
\affiliation{Institut Laue-Langevin, 38000 Grenoble, France}

\author{D.~Lange}
\affiliation{Max-Planck-Institut für Kernphysik, 69117 Heidelberg, Germany}

\author{I.~D.~Moore} 
\affiliation{University of Jyvaskyla, Department of Physics, Accelerator laboratory, P.O. Box 35(YFL) FI-40014 University of Jyvaskyla, Finland}

\author{M.~M\"uller}
\affiliation{Max-Planck-Institut für Kernphysik, 69117 Heidelberg, Germany}

\author{M.~Mougeot}
\affiliation{University of Jyvaskyla, Department of Physics, Accelerator laboratory, P.O. Box 35(YFL) FI-40014 University of Jyvaskyla, Finland}
\affiliation{Max-Planck-Institut für Kernphysik, 69117 Heidelberg, Germany}

\author{D.A.~Nesterenko}
\affiliation{University of Jyvaskyla, Department of Physics, Accelerator laboratory, P.O. Box 35(YFL) FI-40014 University of Jyvaskyla, Finland}

\author{H.~Penttil\"a}
\affiliation{University of Jyvaskyla, Department of Physics, Accelerator laboratory, P.O. Box 35(YFL) FI-40014 University of Jyvaskyla, Finland}

\author{C.~Petrone}
\affiliation{IFIN-HH, P.O. Box MG-6, 077125 Bucharest-Magurele, Romania}

\author{I.~Pohjalainen}
\affiliation{University of Jyvaskyla, Department of Physics, Accelerator laboratory, P.O. Box 35(YFL) FI-40014 University of Jyvaskyla, Finland}

\author{A.~de~Roubin}
\altaffiliation{Present address: KU Leuven, Instituut voor Kern- en Stralingsfysica, B-3001 Leuven, Belgium}
\affiliation{University of Jyvaskyla, Department of Physics, Accelerator laboratory, P.O. Box 35(YFL) FI-40014 University of Jyvaskyla, Finland}

\author{V.~Rubchenya}
\altaffiliation{Deceased.}
\affiliation{University of Jyvaskyla, Department of Physics, Accelerator laboratory, P.O. Box 35(YFL) FI-40014 University of Jyvaskyla, Finland}

\author{Ch.~Schweiger}
\affiliation{Max-Planck-Institut für Kernphysik, 69117 Heidelberg, Germany}

\author{L.~Schweikhard}
\affiliation{Institut für Physik, Universität Greifswald, 17487 Greifswald, Germany}

\author{M.~Vilen}
\affiliation{University of Jyvaskyla, Department of Physics, Accelerator laboratory, P.O. Box 35(YFL) FI-40014 University of Jyvaskyla, Finland}

\author{J.~\"Ayst\"o}
\affiliation{University of Jyvaskyla, Department of Physics, Accelerator laboratory, P.O. Box 35(YFL) FI-40014 University of Jyvaskyla, Finland}

\date{\today}% It is always \today, today,
             %  but any date may be explicitly specified

\begin{abstract}
Isomers close to doubly-magic $^{78}_{28}$Ni$_{50}$ provide essential information on the shell evolution and shape coexistence near the ${Z=28}$ and ${N=50}$ double shell closure.
We report the excitation energy measurement of the $1/2^{+}$ isomer in $^{79}_{30}$Zn$_{49}$ through independent high-precision mass measurements with the JYFLTRAP double Penning trap and with the ISOLTRAP Multi-Reflection Time-of-Flight Mass Spectrometer.
We unambiguously place the $1/2^{+}$ isomer at $\SI{942\pm10}{\kilo\electronvolt}$, slightly below the $5/2^+$ state at $\SI{983\pm 3}{\kilo\electronvolt}$.
With the use of state-of-the-art shell-model diagonalizations, complemented with Discrete Non Orthogonal shell-model calculations which are used here the first time to interpret shape coexistence, we find low-lying deformed intruder states, similar to other ${N=49}$ isotones.
The $1/2^{+}$ isomer is interpreted as the band-head of a low-lying deformed structure akin to a predicted low-lying deformed band in $^{80}$Zn, and points to shape coexistence in $^{79,80}$Zn similar to the one observed in $^{78}$Ni.
The results make a strong case for confirming the claim of shape coexistence in this key region of the nuclear chart.  

\end{abstract}

\maketitle

The atomic nucleus, a conglomerate of protons and neutrons, is a complex many-body system with unique features.
The nuclear shell model has successfully described various nuclear properties, including the emergence of shell closures~\cite{goeppert_mayer1949} and magic numbers~\cite{haxel_jensen_suess1949}.

Similar to the atomic shell, nuclei can be excited, resulting in a dense level structure.
Ground and excited states can show different shapes resulting from the microscopic wave function~\cite{jain2021nuclear}.
Deformed excited states often emerge near closed shells, where the excitation of multiple nucleons across the shell gap can be energetically favorable, leading to deformation through the increased number of particles found in the valence space~\cite{Heyde_Wood_2011}.
While typically the coexistence of ground states and deformed excited states at low energies are observed, shape inversion can appear when the ground state becomes deformed in coexistence with a spherical excited state~\cite{Heyde_Wood_2011, gray2023shapeisomer, Nowacki:2021fjw}

Research on shape coexistence close to the doubly-magic nucleus $^{78}_{28}$Ni$_{50}$ has gained momentum only recently~\cite{Garrett2022}.
Low-lying intruder states, often indicators of shape coexistence, have been studied in ${N=49}$ isotones through transfer-reaction experiments~\cite{ahn2019intruderstates81Ge, montresque1978seintruderstates, detorie1978kr85structure, burks1986Sr87intruderstates}.
First evidence supporting shape coexistence, such as the claimed discovery of a $0^+_2$ intruder state in $^{80}_{32}$Ge$_{48}$ from Ref.~\cite{Gottardo2016}, could, however, not be confirmed in subsequent experiments \cite{Garcia_2020, Sekal2021_IDS_80Ge}.

More recently, the doubly-magic nature of $^{78}$Ni was supported through the measurement of its E($2_1^{+}$) value~\cite{Taniuchi2019}, as well as $\gamma$-ray spectroscopy~\cite{Olivier2017} and mass spectrometry of $^{79}_{29}$Cu$_{50}$~\cite{Welker2017}, reinforcing the persistence of the ${Z=28}$ gap.
The potential appearance of shape coexistence in $^{78}$Ni is furthermore theorized to be a pathway into a new island of inversion at ${N=50}$~\cite{Nowacki2016}.

The isomer in $^{79}_{30}$Zn$_{49}$, a long-lived nuclear state, provides a unique opportunity to further study the interplay between the single-particle and collective degrees of freedom in the close vicinity of $^{78}$Ni.
The first spectroscopy of $^{79}$Zn from (d,p) transfer reactions found evidence for the presence of intruder states and tentatively assigned a spin-parity of $1/2^+$ to the isomeric state with an excitation energy of $\SI{1100\pm150}{\kilo\electronvolt}$, as well as a close-lying $5/2^+$ state with $\SI{983\pm3}{\kilo\electronvolt}$~\cite{Orlandi2015}, leaving the exact state ordering uncertain.
The spin-parity was confirmed for the $9/2^+$ ground and the $1/2^+$ isomeric states through magnetic moment measurements from collinear laser spectroscopy~\cite{Yang2016, Wraith2017}.
More significantly, these works found a large isomer shift in the charge radius but could not connect this increase to a deformation unambiguously.

Recent studies have shown that shape changes may be linked to multiple particle-hole excitations of both protons and neutrons \cite{Morales2017typeIIshellevolution, leoni2017shapeisomerismin66ni, stryjczyk2018decayat66Mnto66Ni, Sels2019, olaizola2019co65properties}.
However, a precise and accurate excitation energy measurement of the $1/2^+$ isomer to confirm claims of shape coexistence in $^{79}$Zn is missing from the above-mentioned work. 
Such measurement will unambiguously determine the state ordering, validate the particle-hole excitation character, and benchmark the binding energy predictions of the employed shell-model interactions near $^{78}$Ni.

While the ground-state mass of $^{79}$Zn is precisely known~\cite{Hakala2008,Baruah2008,Giraud2022} via mass measurements, the uncertainty on the excitation energy from the transfer-reaction experiment is rather large. 
Given the importance of shape coexistence in the immediate vicinity of $^{78}$Ni, we present two independent high-precision mass spectrometry experiments of the isomer $^{79}$Zn$^m$ using the JYFLTRAP double Penning trap~\cite{Eronen2012} at the Ion Guide Isotope Separator On-Line~(IGISOL) facility~\cite{Moore2013} in Jyväskylä (Finland), and the Multi-Reflection Time-of-Flight Mass Spectrometer~(MR-ToF~MS) of ISOLTRAP~\cite{2008_Mukherjee} at ISOLDE/CERN~\cite{2017_Catherall_ISOLDE} (Switzerland).
The experiments were performed at different facilities and with different techniques to ensure the isomer's production and the accuracy of its excitation energy.
The results are interpreted by large-scale shell model calculations, utilizing the valence space of interactions used in Refs.~\cite{Yang2016, Wraith2017, Welker2017, Nowacki2016}.
While offering a more detailed and accurate picture of the nuclear structure in this critical region, the calculations highlight the relative fragility of the doubly-magic shell strength. 

\begin {table*}[t!]
\caption{\label{tab:results} Frequency ratio $r$ or time-of-flight difference $\Delta t$, mass-excess values ME, and excitation energy of the isomer $E$ determined in this work. The values for $J^\pi$ and $T_{1/2}$ are from Ref.~\cite{Yang2016}, ME$_{\text{lit}}$ is deduced from the ground state mass (weighted average of the measurements from Refs.~\cite{Baruah2008, Hakala2008, Giraud2022}) in combination with the excitation energy reported in~\cite{Orlandi2015}.
$^{84}$Kr$^+$ from~\cite{AME2020} was used as a reference for the ToF-ICR measurements, while $^{79}$Zn$^+$ in its ground state from~\cite{Baruah2008, Hakala2008, Giraud2022} was used for the MR-ToF MS measurements.
}

\begin{ruledtabular}
\begin{tabular}{lcccccccc}
Nuclide  & $J^\pi$ & $T_{1/2}$ & Method & $r$ or $\Delta t$ & ME $(\si{\kilo\electronvolt})$ & ME$_{\text{lit}}$ $(\si{\kilo\electronvolt})$ & Diff. $(\si{\kilo\electronvolt})$ & $E$ $(\si{\kilo\electronvolt})$  \\
\hline \\ [-8pt]
$^{79}$Zn$^m$   & $1/2^+$ & $>200$ ms & \makecell{ToF-ICR\\MR-ToF MS} &\makecell{$0.940796186(144)$\\$\SI{274.2\pm 4}{\nano\second}$} & \makecell{$-52490(12)$\\$-52489(14)$} & $-52332(150)$ & \makecell{$-158(150)$\\$-157(150)$} & \makecell{$942(12)$\\$943(14)$} \\
\end{tabular}
\end{ruledtabular}
\end{table*}

\begin{figure}[t]
	\centering
	\includegraphics[width=\columnwidth]{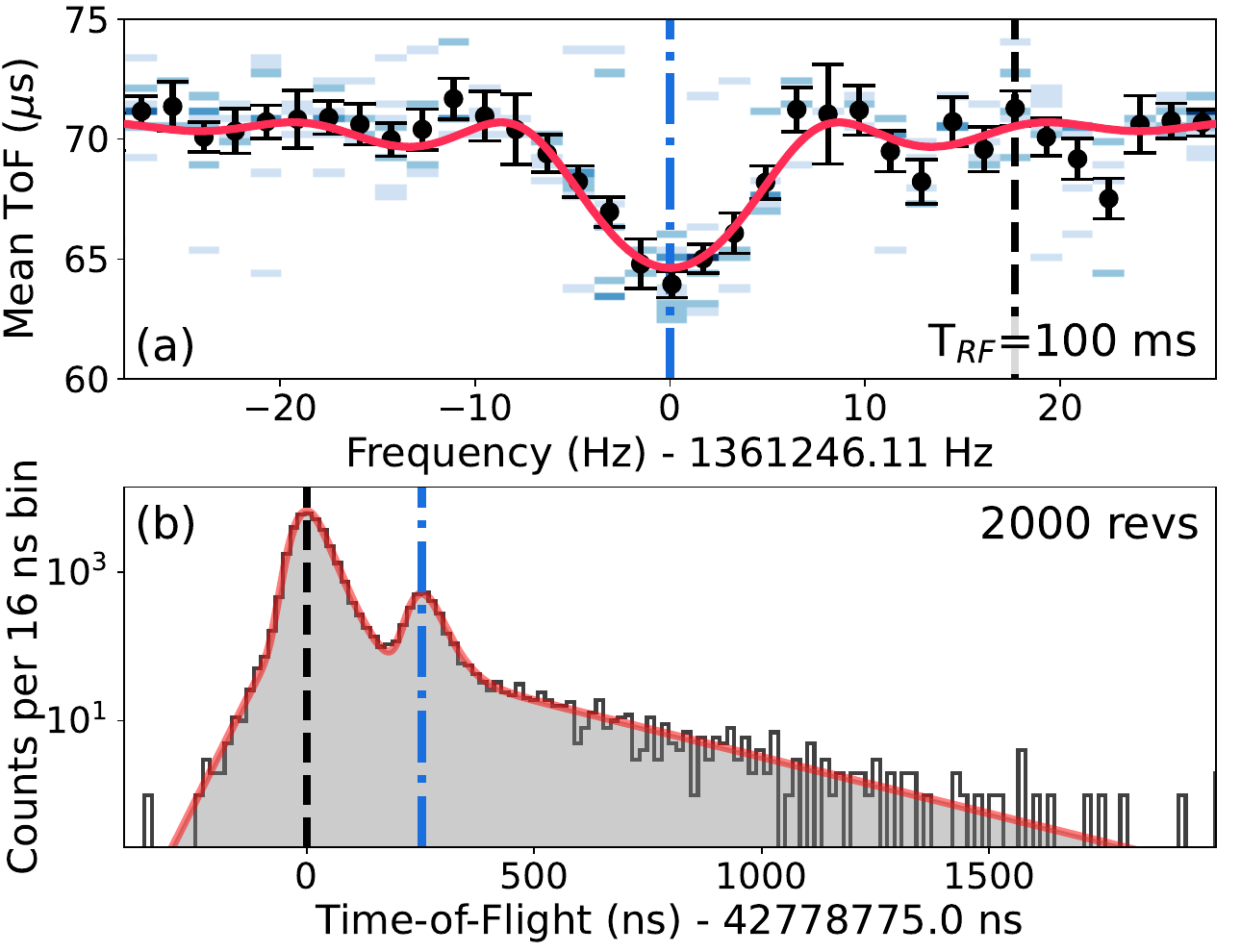} 
\caption{(a) A typical ToF-ICR spectrum for the $1/2^+$ state in $^{79}$Zn$^+$.
Colored bins indicate the number of detected ions. Darker shades correspond to more ions and lighter shades to fewer ions.
The solid red line represents the fit to the data points (black) using the model from Ref.~\cite{Konig1995}.
The cyclotron frequencies are indicated with a vertical black dashed line for the ground state (not present in this spectrum) and a vertical blue dash-dotted line for the isomer.
(b) Time-of-Flight spectrum for the MR-ToF MS data.
The ToF of the ground state is indicated by a vertical black dashed line, and the ToF of the isomer by a vertical blue dash-dotted line. The solid red line represents the fit to the data using the model from Ref.~\cite{2017_Purushothaman_hyperEMG}.}
\label{fig:measurements}
\end{figure}

\textit{For the Penning-trap measurements}, the ions of interest were produced via proton-induced fission using \mbox{$35$-MeV} protons from the K130 cyclotron impinging onto a $\SI{15}{\milli\gram\per\square\centi\meter}$ thick $^{nat}$U target at IGISOL.
The reaction products were stopped and thermalized in the helium gas cell of the fission ion guide~\cite{AlAdili2015}, leaving a large fraction of the products singly charged.
The ions were extracted from the chamber with a sextupole ion guide~\cite{karvonen2008spig} and accelerated to 30 keV.
A 55$^\circ$ dipole magnet was used to mass-separate the ions based on their mass-to-charge ratio $m/q$.
The mass-separated beam was then stopped in a radiofrequency quadrupole cooler and buncher (\mbox{RFQ-cb})~\cite{Nieminen01} and released as ion bunches into the double Penning trap JYFLTRAP~\cite{Eronen2012}.
In the first trap, either the ground- or isomeric-state ions of $^{79}$Zn were selected using mass-selective buffer-gas cooling~\cite{Savard1991}.
The selected ions were transferred to the second trap, where the high-precision mass measurements were performed using the Time-of-Flight Ion Cyclotron Resonance~(\mbox{ToF-ICR}) method~\cite{Konig1995}.
The cyclotron resonance frequency ${\nu_c=qB/(2\pi m)}$ of the $1/2^+$ state in $^{79}$Zn$^{+}$ was measured using a $\SI{100}{\milli\second}$ pulse of quadrupolar rf excitation.
Altogether four ToF-ICR spectra were measured for the $1/2^+$ state in $^{79}$Zn$^{+}$. 
An example of such a ToF-ICR spectrum is shown in Fig.~\ref{fig:measurements}(a).

The magnetic field strength $B$ was determined using $^{84}$Kr$^+$ as a mass reference.
The mass of the $1/2^+$ state in $^{79}$Zn was obtained from the measured frequency ratio $r=\nu_{c,ref.}/\nu_c$ between the $^{84}$Kr$^+$ reference ions and the isomeric-state ions of $^{79}$Zn$^+$ as $m=r(m_{ref}-m_e)+m_e$, where $m_{ref}$ is the mass of the reference ion and $m_e$ is the electron mass.
Two sources of systematic uncertainties were taken into account in the analysis, the fluctuation of the magnetic field being $8.18 \times 10^{-12} \times \Delta t \mathrm{~min}^{-1}$~\cite{Canete2016}, where $\Delta t$ represents the time between two reference measurements, and mass-dependent uncertainties being $2.2 \times 10^{-10} \times (m - m_{ref})/\mathrm{u}$~\cite{Canete2019}. 

\textit{For the MR-ToF MS measurements}, neutron-rich zinc isotopes were produced at the isotope separation online facility ISOLDE at CERN~\cite{2017_Catherall_ISOLDE} by impinging a \mbox{$1.4$-GeV} proton beam onto a solid tungsten block to generate an intense spallation neutron flux~\cite{Luis2012}. 
The zinc isotopes were then produced through neutron-induced fission processes in an adjacent thick uranium carbide target.
Using the tungsten converter resulted in reduced production of isobaric neutron-deficient nuclides.
The radioactive fission products then diffused through the target material into a cold quartz transfer line~\cite{Bouquerel2007, Bouquerel2008} which further eliminated contamination of surface-ionized elements, e.g. Ga, Rb, and Sr~\cite{Koester2008}.
The remaining radioactive species were then ionized by the Resonance Ionization Laser Ion Source (RILIS)~\cite{2017_Fedosseev_RILIS}, using an element-selective three-step ionization scheme for zinc. 
The ion beam was then mass-separated using the general-purpose mass-separator dipole magnet, removing non-isobaric contamination, before being sent to the ISOLTRAP mass spectrometer~\cite{2008_Mukherjee}.

The quasi-continuous ion beam was cooled and bunched in a linear RFQ-cb~\cite{HERFURTH2001254} with a storage time of $\SI{10}{\milli\second}$ before being captured in the MR-ToF MS~\cite{2013_wolf} using the in-trap lift method~\cite{wolf2013staticmirrorioncapture}. 
After trapping times of up to $\SI{43}{\milli\second}$, the ion bunch was ejected onto a single-ion counting detector using the same in-trap lift. 

The excitation energy of the isomeric state ${E=\left[(\Delta t/t_0)^2+2\Delta t/t_0\right]m_0c^2}$ is related to the time-of-flight difference $\Delta t$ between the ground state with mass $m_0$ and the isomeric state, the absolute flight time of the ground state $t_0$, and the speed of light in vacuum $c$.
For long flight times, where ${\Delta t \ll t_0}$, this reduces to ${E\approx2\Delta t/t_0\times m_0c^2}$.
Figure~\ref{fig:measurements}~(b) shows the ToF spectrum for the MR-ToF~MS data. 
Zinc was delivered from ISOLDE as a pure beam, thus only the ground state (black dashed line) and the isomeric state (blue dash-dotted line) of $^{79}$Zn were present in the spectrum.
A mass resolving power $R=t_0/2\Delta t_\text{FWHM}=300\,000$ was reached, sufficient to resolve the two states.
The asymmetric peak shape was fitted with a multi-component exponentially modified Gaussian (``hyper-EMG'')~\cite{2017_Purushothaman_hyperEMG} to extract the absolute ToF $t_0=\SI{4277871\pm2}{\nano\second}$ of the ground state and the ToF difference $\Delta t=\SI{274\pm 4}{\nano\second}$ between the two states.
The excitation energy was then calculated using the ground state mass with ME$_{\text{lit}}=\SI{-53432.1\pm1.8}{\kilo\electronvolt}$, which we calculated as the weighted mean of the results from Refs.~\cite{Hakala2008,Baruah2008,Giraud2022}.
The excitation energy was measured for different ion loads to account for ion-ion interactions during the storage time in the MR-ToF MS~\cite{rosenbusch2013spacecharge, Maier2023spacecharge}.

The results of the two independent measurements are summarized in Tab.~\ref{tab:results}.
The extracted excitation energies of the isomer agree very well, resulting in a weighted mean of $\SI{942\pm10}{\kilo\electronvolt}$.
The isomer energy is lower than $\SI{1100\pm150}{\kilo\electronvolt}$ as given in the NUBASE 2020 evaluation~\cite{Kondev2021}, which is based on the transfer reaction experiment from Ref.~\cite{Orlandi2015}.
Our value is significantly more precise and unambiguously sets the isomeric state below the $5/2^+$ state located at $\SI{983\pm3}{\kilo\electronvolt}$.
We note that the result agrees with the value $\SI{943\pm3}{\kilo\electronvolt}$, obtained from the beta-decay spectroscopy of $^{79}$Cu~\cite{Delattre2016}.
Here, we confirm the existence of the isomer and provide a direct measure of its excitation energy.

\begin {table*}[tbp]
\caption{\label{tab:theo} Occupation of orbitals in the full proton $pf$ and neutron $sdg$ valence space for low-lying states in $^{79,80}$Zn and $^{78}$Ni (the latter taken from Ref.~\cite{Nowacki2016}). E$_{\rm{exp}
}$ and E$_{\rm{theo}}$ (in $\si{\mega\electronvolt}$) are the experimental and theoretical excitation energies. E$_{\rm{corr}}$ and E$^*_{\rm{corr}}$ (in MeV) are the total correlation energy and the correlation energy difference of an excited state with respect to its ground state. $n^*_\nu$ and $n^*_\pi$ are the total numbers of protons and neutrons above Z=28 and N=50, respectively.
}
\begin{ruledtabular}
\begin{tabular}{lc|cc|cc|cccccc|ccccc}
Nuclide  & $J^\pi$ & E$_{\rm{exp}}$ & E$_{\rm{theo}}$ & E$_{\rm{corr}}$ & E$^*_{\rm{corr}}$ & $n^*_\nu$ & $\nu_{g9/2}$ &  $\nu_{d5/2}$ &  $\nu_{s1/2}$  & $\nu_{g7/2}$ &  $\nu_{d3/2}$ & $n^*_\pi$ & $\pi_{f7/2}$ &  $\pi_{f5/2}$  & $\pi_{p3/2}$ &  $\pi_{p1/2}$  \\ %[5pt] 
\hline
$^{79}$Zn 		& $9/2^+$ & 0.0  & 0.0  & -11.72 & - & 0.53 & 8.47  & 0.27 & 0.04 & 0.18 & 0.04 & 2.49 & 7.51 & 1.79 & 0.50 & 0.20 \\
                & $1/2^+$ & 0.94 & 0.83 & -18.59 & -6.87  & 1.84 &  7.17  & 0.81 & 0.54 & 0.34 & 0.15 & 2.82 & 7.18 & 1.45 & 0.95 & 0.42 \\
                & $5/2^+$ & 0.98 & 0.94 & -18.23 & -6.51 & 1.82 & 7.18  & 1.06 & 0.31 & 0.33 & 0.12 & 2.79 &  7.20 & 1.51 & 0.87 & 0.41 \\ [5pt]    
$^{80}$Zn 		& $0^+_1$ & 0.0  & 0.0 & -10.80 & - & 0.49  & 9.50  & 0.23 & 0.03 & 0.19 & 0.04 & 2.48 & 7.52 & 1.90 & 0.44 & 0.14 \\
                & $0^+_2$ & -  & 2.16 & -17.12 & -6.32  & 2.74  & 7.26  & 1.20 & 0.71 & 0.52 & 0.31 & 3.08 &  6.92 & 1.33 & 1.28 & 0.47 \\[5pt]
$^{78}$Ni 		& $0^+_1$ & 0.0  & 0.0 & -8.00 & - & 0.38  & 9.62  & 0.12 & 0.02 & 0.20 & 0.04 & 0.57 & 7.44 & 0.38 & 0.15 & 0.04 \\
                & $0^+_2$ & -  & 2.65 & -24.09 & -16.09  & 2.70  & 7.30  & 1.11 & 0.81 & 0.43 & 0.35 & 2.35 &  5.65 & 0.98 & 0.94 & 0.43 \\%[1pt]               
\end{tabular}
\end{ruledtabular}
\end{table*}

To interpret the present experimental data, shell-model calculations with the \mbox{PFSDG-U} interaction~\cite{Nowacki2016,Taniuchi2019} were performed for $^{79,80}$Zn.
The valence space is spanned across the full $pf$ shell for protons and full $sdg$ shell for neutrons, with $^{60}$Ca as an inert core.
This interaction has been successfully used in the $^{78}$Ni region to describe, among others, the two-neutron separation energies $S_{2n}$ along the zinc isotopic chain~\cite{Welker2017}, as well as the magnetic g-factor in $^{79}$Zn~\cite{Wraith2017}.

The calculated excitation energies for $^{79}$Zn (Tab.~\ref{tab:theo}) are in good agreement with the experimental results with $1/2^+$ and $5/2^+$ at $\SI{0.83}{\mega\electronvolt}$ and $\SI{0.94}{\mega\electronvolt}$, respectively.
We find that the two low-lying excited states in $^{79}$Zn show a one-particle-two-hole configuration, consistent with other ${N=49}$ isotones~\cite{ahn2019intruderstates81Ge, montresque1978seintruderstates, detorie1978kr85structure, burks1986Sr87intruderstates}.
While the s$_{1/2}$ and d$_{5/2}$ orbitals lie together at the neutron Fermi surface in the vicinity of $^{78}$Ni, the correlated ${N=50}$ neutron gap ${S_N(^{81}\text{Zn}) - S_N(^{80}\text{Zn})}$, calculated from the present theoretical values, remains sizeable at about $\SI{4.0}{\mega\electronvolt}$ and is in agreement with the effective single-particle energies of Ref.~\cite{Nowacki:2021fjw} and with the experimental value provided in Ref.~\cite{AME2020}. 
These states usually recover enough correlation energy (total energy minus the monopole part) to compensate for their energy gap loss.
This is observed in our calculations shown in Tab.~\ref{tab:theo}, where the total correlation energy is extracted.
The excited $1/2^+$ and $5/2^+$ states recover correlation energy on the order of $\sim$ $\SI{6.5}{\mega\electronvolt}-\SI{6.9}{\mega\electronvolt}$ compared to the $9/2^+$ the ground state.

The low excitation energy of these two states can be understood as the balance between an average of 1.3 neutrons excited across the shell gap with respect to the ground state ({$\sim\SI{5.5}{\mega\electronvolt}$}) and the correlation energies ({$\sim\SI{6.5}{\mega\electronvolt}-\SI{6.9}{\mega\electronvolt}$}) which compensate and result in very low-lying excitation energies for these one-particle-two-holes states.

\begin{figure}[t]
\centering
\includegraphics[width=1\columnwidth]{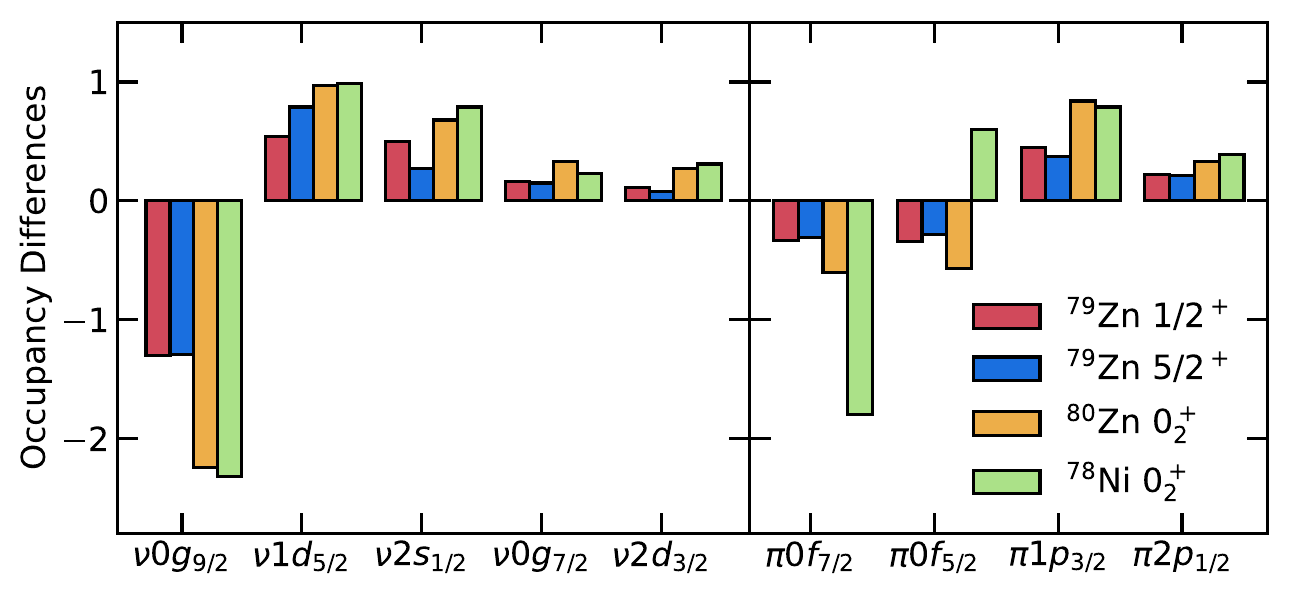}
\caption{Occupancy differences between excited states in $^{79,80}$Zn, and $^{78}$Ni with their respective ground states. The data for $^{78}$Ni is taken from~\cite{Nowacki2016}.}
\label{fig:occupancies_single}
\end{figure}

An inspection of the partial occupancies of the first two excited states (see Tab.~\ref{tab:theo}) reveals strong neutron mixing for the orbitals above ${N=50}$, as well as different proton occupancies with respect to the ground state, invalidating the spherical single particle-hole nature of these states and rather suggesting a deformed shape.
This neutron mixing and proton reshuffling is visualized in Fig.~\ref{fig:occupancies_single}, where the occupancy differences of the excited states with respect to their ground states are plotted. 

The structure of the excited states can be interpreted within the Discrete Non-Orthogonal shell-model (DNO-SM) method, newly developed in Ref.~\cite{Dao:2022igs} and applied recently in Refs.~\cite{Rocchini:2023scg,Rezynkina:2022txa}.
This approach expands the shell-model wave functions in the deformed Hartree-Fock basis rather than the usual spherical m-scheme basis.
This allows the extraction of the corresponding deformation amplitudes of a given state in the $(\beta,\gamma)$ plane.

\begin{figure}[t]
\centering
\includegraphics[width=1.0\columnwidth]{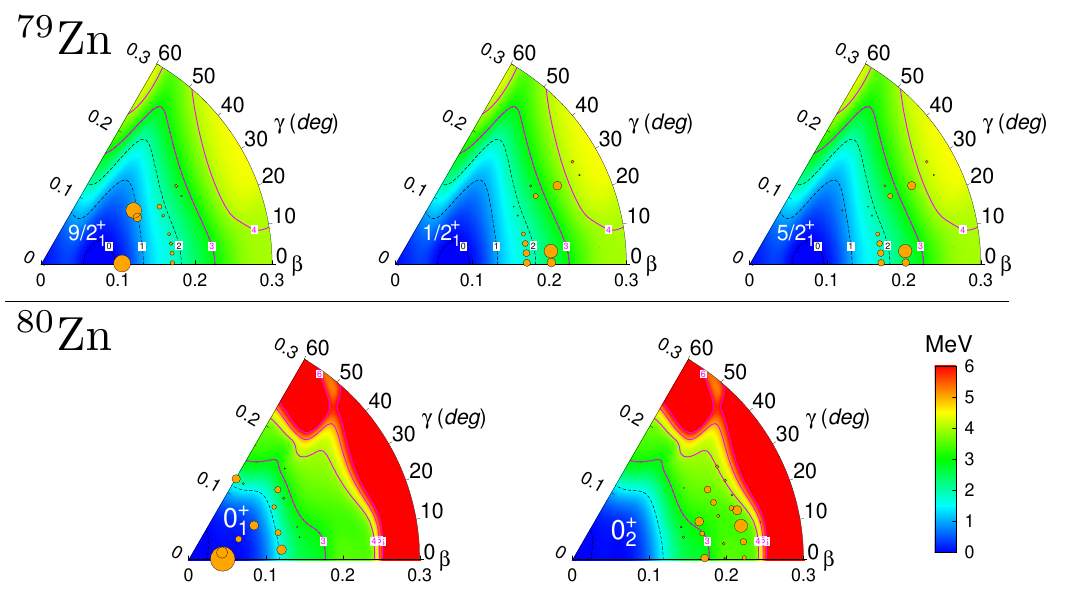}
\caption{DNO-SM expansions in the ($\beta, \gamma$) plane (using the same energy scale) for low-lying states in $^{79}$Zn ($9/2^+_1$, $1/2^+_1$ and $5/2^+_1$ in upper panel) and $^{80}$Zn ($0^+_1$ and $0^+_2$ in lower panel).
The radius of circles represents the normalized probability of finding a deformation of ($\beta, \gamma$) in the corresponding state.} 
\label{PESZn79-80}
\end{figure}

Figure~\ref{PESZn79-80} (top panel) depicts such expansion for the ground state and the first two excited states of $^{79}$Zn.
Two clear patterns emerge: in the ground state, the main components of the wave function tend towards the low-deformation region with a small $\beta$ value, while for the excited states, the wave functions are more fragmented and have, on average, a larger deformation.
Also, we find the clustering of the wave-function components consistent with the deformation parameters $\beta\approx0.15$ (ground state) and $\beta\approx 0.22$ (isomeric state), deduced from the nuclear charge radius in Ref.~\cite{Yang2016}.
Moreover, the DNO-SM calculations reveal that both $1/2^+$ and $5/2^+$ states belong to the same rotational structure, which is characterized by the $K=1/2$ components of 100\% and 96\%, respectively (see K-component extraction in Refs.~\cite{Dao:2022igs,Rezynkina:2022txa}).

To probe further into the deformed character of the low-lying states in $^{79}$Zn, we have complemented the calculations with the investigation of $^{80}$Zn.
We find two low-lying $0^+$ states: the spherical ground-state and an excited deformed state of two-particle-two-hole nature at $\SI{2.16}{\mega\electronvolt}$ (see Tab.~\ref{tab:theo}).
Again, a larger correlation energy ($\sim$ $\SI{6.3}{\mega\electronvolt}$) is observed for the excited 0$^+$ state on the same order of magnitude as those of the deformed intruder states in $^{79}$Zn, indicating the same deformation nature of these states.
Figure~\ref{PESZn79-80} shows the wave function expansions for these two 0$^+$ states (bottom panel).
There is a clear similarity between the spherical ground states for \mbox{$^{79}$Zn($9/2^+$)/$^{80}$Zn($0^+_1$)} and the deformed excited states for \mbox{$^{79}$Zn($1/2^+$, $5/2^+$)/$^{80}$Zn($0^+_2$)}, advocating for the deformed nature of the observed isomer in $^{79}$Zn, as well as its $5/2^+$ companion.

Finally, the present shape coexistence discussed for $^{79,80}$Zn can be put in perspective with the shape coexistence recently observed and discussed for $^{78}$Ni~\cite{Nowacki2016,Taniuchi2019}: the deformed intruder 0$^+_2$ of $^{78}$Ni has 2.7 neutron p-h excitations on average, in remarkable agreement with present values quoted in Tab.~\ref{tab:theo} for 0$^+_2$ of $^{80}$Zn.
Both 0$^+_2$ states have $\sim2.4-3$ protons on average in the ${f_{5/2}, p_{3/2}, p_{1/2}}$ shells, leading to close collective structures.
Therefore, the shape coexistence in $^{78}$Ni and in the presented $^{79-80}$Zn reveal striking similarities.

To summarize, we have established the level ordering and determined the excitation energy of the isomer in $^{79}$Zn by means of high-precision mass spectrometry.
Two measurements were performed independently, using different production methods and measurement techniques at different radioactive ion beam facilities.
We show unambiguously that the $1/2^+$ isomeric state with $\SI{942\pm10}{\kilo\electronvolt}$ lies below the $5/2^+$ state with $\SI{983\pm3}{\kilo\electronvolt}$.
The new DNO-SM calculations tool provides the theoretical analysis, predicting the occurrence of low-lying deformed intruder states.
The $1/2^+$ isomer is interpreted as the band-head of a low-lying deformed structure of the same nature as a predicted low-lying deformed band in $^{80}$Zn.
These findings provide an additional indication for shape coexistence in $^{79,80}$Zn, similar to the one suggested for $^{78}$Ni.

We thank the ISOLDE technical group and the ISOLDE Collaboration for their support.
We acknowledge the support of the German Max Planck Society, the French Institut National de Physique Nucléaire et de Physique des Particules (IN2P3), the European Research Council (ERC) under the European Union’s Horizon 2020 research and innovation programme (Grant Agreements No.~682841 ‘ASTRUm’, 654002 ‘ENSAR2’, and 101020842 ‘EUSTRONG’), as well as the German Federal Ministry of Education and Research (BMBF; Grants No.~05P18HGCIA, 05P21HGCI1, and 05P21RDFNB).
L.N. acknowledges support from the Wolfgang Gentner Programme of the German Federal Ministry of Education and Research (Grant No.~13E18CHA).
This work has been supported by the Academy of Finland under the Finnish Centre of Excellence Program (Nuclear and Accelerator Based Physics Research at JYFL 2012-2017), and under Academy of Finland grants No. 275389, 284516, 312544, 295207, and 306980. We acknowledge the support of the European Union's Horizon 2020 research and innovation program grant agreement No 654002 (ENSAR2) and No. 771036 (ERC CoG MAIDEN). We thank for the bilateral mobility grant from the Institut Fran$\text{\c{c}}$ais in Finland, the Embassy of France in Finland, the French Ministry of Higher Education and Research, and the Finnish Society of Science and Letters. We are grateful for the mobility support from PICS MITICANS (Manipulation of Ions in Traps and Ion sourCes for Atomic and Nuclear Spectroscopy). S.G. thanks for the mobility grant from the EDPSIME. F. N and D. D. D. acknowledge the financial support of CNRS/IN2P3, France, via ABI-CONFI master projet.

The JYFLTRAP experiment was conducted by L.C, S.G, A.K, B.B, P.A, T.E, V.G.A., A.J., A.K, I.D.M, D.A.N, F.D.O., H.P., C.P., I.P., A.D.R., V.R., M.V., and J.Ä.
The ISOLTRAP experiment was conducted by L.N., R.B.C., P.F., M.F., A.H., D.La., M.Mü., M.M., Ch.S., and was conceived by U.K.  
The theoretical calculations were performed by D.D.D. and F.N. Funding and supervision were provided, in parts, by K.B. and L.S.
The Manuscript was prepared by L.N., D.D.D., A.K., D.Lu., F.N., and M.S.
All authors contributed to the editing of the manuscript. 

\bibliography{bib}

\end{document}